\documentstyle[a4,12pt,amssymb,times]{article}

\newcommand{\be}{\begin{equation}}
\newcommand{\ee}{\end{equation}}
\newcommand{\bea}{\begin{eqnarray}}
\newcommand{\eea}{\end{eqnarray}}

\def\p1{\pi_1}

\def\l{\lambda}
\def\f{\phi}

\def\r{\rho}

\def\pmas{\partial_+}
\def\pmen{\partial_-}

\begin{document}

\vspace*{\stretch{0}}
\begin{flushright}
{\tt FTUV-00-1204A\\
     IFIC/00-78\\
     hep-th/0012019}
\end{flushright}

\vspace*{0.5cm}

\begin{center}
{\Large \bf Evaporation of charged black holes near extremality\footnote{Talk
given at the Ninth Marcel Grossmann Meeting.}}
\\[0.5cm]
A. Fabbri$^{\rm a}$\footnote{\tt fabbria@bo.infn.it},
D. J. Navarro$^{\rm b}$\footnote{\tt dnavarro@ific.uv.es} and
J. Navarro-Salas$^{\rm b}$\footnote{\tt jnavarro@ific.uv.es}
\\[0.5cm]
{\footnotesize
a) Dipartimento di Fisica dell'Universit\`a di Bologna and INFN
sezione di Bologna,\\
Via Irnerio 46, 40126 Bologna, Italy.
\\[0.5cm]
b) Departamento de F\'{\i}sica Te\'orica and IFIC, Centro Mixto Universidad
de Valencia-CSIC.\\
Facultad de F\'{\i}sica, Universidad de Valencia, Burjassot-46100, Valencia,
Spain.}
\end{center}

The quantum evolution of black holes including back reaction effects is an open
problem. One can simplify it by considering the scattering of low-energy
neutral particles off extremal Reissner-Nordstr\"om black holes \cite{stlo}.
If, in addition, we also consider spherically reduced configurations, the
theory can be described, in a region very close to the horizon, by a solvable
effective model. This model is just the Jackiw-Teitelboim model \cite{jt}
(for details see \cite{fnn})
\be
\label{effective}
I = \int d^2x \sqrt{-g} \left[ (R + \frac{4}{l^2q^3}) \tilde{\f} - \frac{1}{2}
\sum_i^N |\nabla f_i|^2 \right] \, ,
\ee
where the two-dimensional fields of (\ref{effective}) are related to the
four-dimensional metric by the expression
$ds_{(4)}^2 = \frac{2l}{r_0} ds_{(2)}^2 + (r_0^2+4l^2\tilde{\f}) d\Omega^2$,
$r_0=lq$ is the extremal radius, and $l^2=G$ is Newton's constant. The
fields $f_i$ represent null matter. To analyze back reaction effects we have to
add to the classical action (\ref{effective}) the Polyakov-Liouville term
\be
\label{polyakov}
I_{PL} = -\frac{N\hbar}{96\pi} \int d^2x \sqrt{-g} R \; \square^{-1} R +
\frac{N\hbar}{12\pi} \int d^2x \sqrt{-g} \lambda^2 \, .
\ee
The above expression contains a cosmological term with $\l^2=l^{-2}q^{-3}$ to
ensure that the extremal geometry remains an exact solution of the one-loop
theory. The equations of motion derived from $I+I_{PL}$ in conformal gauge
$ds^2=-e^{2\r}dx^+dx^-$ are
\bea
\label{eqs}
&2\pmas \pmen \r + \l^2 e^{2\r} = 0 \, , \quad \pmas \pmen \tilde{\f} + 
\l^2 \tilde{\f} e^{2\r} = 0 \, , \quad \pmas \pmen f_i = 0 \, ,& \\
\label{const}
&-2\partial^2_{\pm} \tilde{\f} + 4 \partial_{\pm} \rho \partial_{\pm} 
\tilde{\f} = T^f_{\pm \pm} - \frac{N\hbar}{12\pi} t_{\pm} -
\frac{N\hbar}{12\pi} \left( (\partial_{\pm} \r )^2 -
\partial_{\pm}^2 \r \right) \, .&
\eea
The first equation implies that the two-dimensional metric has a negative
constant curvature. To find solutions to the above equations it is convenient
to choose the following form of the metric
\be
ds^2 = - \frac{2l^2q^3 dx^+ dx^-}{(x^--x^+)^2} \, .
\ee
This way only the $t_{\pm}$ terms survive in the quantum correction to the
equations of motion. However it is the choice of those functions $t_{\pm}$ the
crucial point to get the physical solutions. If $u$ and $v$ are the null
coordinates associated to the Eddington-Finkelstein coordinates of the
Reissner-Nordstr\"om metric, the absence of ingoing and outgoing fluxes at
extremality (where $u=x^-$ and $v=x^+$) is given by the conditions
$t_{vv}(v)=0=t_{uu}(u)$. However, once we send matter the relation $v=v(x^+)$
is not the identity although the coordinate $u$ remains equal to $x^-$. This
means that in the general situation we have that $t_(x^-)=0$ and $t_+(x^+) =
\frac{1}{2} \{ v, x^+ \}$. The incoming quantum flux will be given by
$\langle T^f_{vv}(v)\rangle = -\frac{N\hbar}{12 \pi} \left[ (\partial_v \r)^2
- \partial^2_v \r \right]$, which turns out to be
\be
\langle T^f_{vv}(v) \rangle = -\frac{N\hbar}{12 \pi} \{ v, x^+ \}
\left( \frac{dx^+}{dv} \right)^2  \, .
\ee
The equations of motion can be solved immediately in the Vadya-type metric
\be
\label{rnevaporating}
ds^2 = -\left( \frac{2\tilde{x}^2}{l^2q^3}-l\tilde{m}(v) \right) dv^2 +
2d\tilde{x} dv \, , \quad \tilde{\f} = \frac{\tilde{x}}{l} \, ,
\ee
where $\frac{d\tilde{m}(v)}{dv} = \langle T^f_{vv}(v) \rangle + T^f_{vv}$. The
problem is to find the relation between $x^+$ and $v$ coordinates. Defining
$F(x^+)$ as $F=lq^3\frac{dx^+}{dv}$ and jumping from (\ref{rnevaporating}) to
the conformal gauge we find that
\be
\label{evphi}
\tilde{\phi} = \frac{F(x^+)}{x^--x^+} + \frac{1}{2}  F'(x^+) \, .
\ee
The equations of motion, in conformal gauge, imply the following differential
equation for $F$
\be
\label{diffequation}
F'''= \frac{N\hbar}{24\pi} \left( -\frac{F''}{F} + \frac{1}{2} \left(
\frac{F'}{F} \right)^2 \right) - T^f_{++}(x^+) \, .
\ee
In terms of the mass function the evolution equation is then
\be
\label{masseq}
\partial_v \tilde{m}(v) = -\frac{N\hbar}{24\pi lq^3} \tilde{m}(v) +
T^f_{vv} \, ,
\ee
and, at late times, $m(v)$ is given by
$\tilde{m}(v)=\Delta m e^{-\frac{N\hbar}{24\pi lq^3}(v-v_0)}$.
This implies that as time passes the evaporating black hole approaches the
extremal configuration. Moreover the equation (\ref{masseq}) suggests that the
incoming (negative) quantum flux contains the information of the classical
stress tensor $T^f_{vv}$.

\end{document}